\documentclass[12pt]{article}
\usepackage{graphicx}
\usepackage{amssymb}
\usepackage{amsmath}

\newcommand{\bear}{\begin{array}}  
\newcommand {\eear}{\end{array}}
\newcommand{\bea}{\begin{eqnarray}}   
\newcommand{\eea}{\end{eqnarray}}
\newcommand{\beq}{\begin{equation}}   
\newcommand{\eeq}{\end{equation}}
\newcommand{\bef}{\begin{figure}}  \newcommand 
{\eef}{\end{figure}}
\newcommand{\bec}{\begin{center}}  \newcommand 
{\eec}{\end{center}}
  
\newcommand{\la}{\left\langle}  
\newcommand{\ra}{\right\rangle}

\def\EQ#1{Eq.~(\ref{#1})}

\def\lrf#1#2{ \left(\frac{#1}{#2}\right)}
\def\lrfp#1#2#3{ \left(\frac{#1}{#2} 
\right)^{#3}}
\def\GEV#1{10^{#1}{\rm\,GeV}}

\begin{document}

\begin{titlepage}

\begin{flushright}
IPMU 09-0082 \\
ICRR-Report-549
\end{flushright}

\vskip 1.35cm

\begin{center}
{\large \bf
The R-axion and non-Gaussianity
 }
\vskip 1.2cm

Kazunori Nakayama$^a$
and
Fuminobu Takahashi$^b$

\vskip 0.4cm

{ \it $^a$Institute for Cosmic Ray Research,
University of Tokyo, Kashiwa 277-8582, Japan}\\
{\it $^b$Institute for the Physics and Mathematics of the Universe,
University of Tokyo, Kashiwa 277-8568, Japan}
\date{\today}

\begin{abstract}
We study cosmological implications of  an R-axion, a pseudo Nambu-Goldstone boson associated with  spontaneous
breaking of an $U(1)_R$ symmetry,  focusing on its quantum fluctuations generated 
during inflation.  We show that, in the anomaly mediation,  
the R-axion decays into a pair of gravitinos, which eventually decay into the visible particles
producing the neutralino LSP. As a result, the quantum fluctuations of the R-axion are
inherited by the cold dark matter isocurvature density perturbation with potentially large non-Gaussianity. The constraints
on the inflation scale and the initial misalignment are derived.
\end{abstract}


\end{center}
\end{titlepage}

\section{Introduction} \label{intro}

Supersymmetry (SUSY) stabilizes the weak scale against radiative corrections
and  offers a solution to the gauge hierarchy problem.  To explain the smallness of the weak scale 
compared to the ultraviolet  scale such as the Planck scale, however,
SUSY must be broken dynamically. It is well known that constructing 
a viable dynamical SUSY breaking (DSB) model is a highly non-trivial task.  In particular, it was shown that, in 
calculable models with a generic superpotential, the existence of an R symmetry is a necessary condition for having a SUSY breaking vacuum
and that a spontaneously broken R symmetry is a sufficient condition~\cite{Nelson:1993nf}.

The spontaneously broken continuous R symmetry leads to a Nambu-Goldstrone boson, 
the so called R-axion. The R-axion acquires a  mass from an R-symmetry breaking.
In particular,  the continuous R symmetry must be explicitly broken by a constant term in a superpotential, $W = C_0$, 
in order for the cosmological constant to vanish. Therefore the R-axion becomes
necessarily massive.  

The cosmology of the R-axion was studied in Ref.~\cite{Bagger:1994hh}, assuming that 
the constant $C_0$ is the main source of the R-axion mass. The initial misalignment of the R-axion from
the potential minimum gives rise to coherent oscillations, which may affect the standard cosmology such as
the big bang nucleosynthesis (BBN). Their main conclusion was that, 
in renormalizable hidden sector models where the SUSY breaking in a DSB sector is transmitted to
the visible sector via Planck-suppressed operators, the R-axion is cosmologically harmless 
since it can be as heavy as ${\cal O}(10^7)$\,GeV and decays before BBN. It was also pointed out
that the R-axion decay may produce a large amount of the gravitinos, which in turn sets an upper bound
on the reheating temperature to avoid conflicts with BBN.

In this letter we study the cosmology of the R-axion,  focusing on its quantum 
fluctuations generated during inflation. 
As discussed recently in Ref.~\cite{Kawasaki:2008sn},
a light scalar such as the R-axion may generate sizable
non-Gaussianity in the isocurvature  fluctuation, as in the case of the QCD axion.
As we will see in the following sections,  the quantum fluctuations of the R-axion are inherited
by the neutralino dark matter (DM) non-thermaly produced by the gravitino decay.
A sizable fraction of the parameter space in the inflation scale 
and the initial misalignment turns out to be already excluded by the current observations. 
The isocurvature perturbation and its associated non-Gaussianity, therefore, 
put a tight constraint on the R-axion cosmology.

\section{R-axion mass and abundance} \label{sec2}

We consider the dynamical SUSY breaking in which SUSY is broken
by the strong gauge dynamics in the SUSY breaking sector. Let us denote
the dynamical scale by $\Lambda$. We also assume that the  R  symmetry
is spontaneously broken as a result of the SUSY breaking, and $f$ denotes
the breaking scale of the R symmetry. Then the R-axion arises as a Nambu-Goldstone boson.
It is known that the constant term in the superpotential,
which is needed to cancel the cosmological constant, explicitly breaks the
$U(1)_R$ symmetry, generating a non-vanishing mass of the R-axion.
Assuming that the constant term is the dominant source of the
explicit  R -breaking, we can estimate the R-axion mass as~\cite{Bagger:1994hh}
\beq
m_a^2 \;\simeq\;\frac{m_{3/2}^2M_P}{f} \simeq 2 \times 10^{16} {\rm GeV}^2 
\lrfp{m_{3/2}}{100{\rm\,TeV}}{2} \lrfp{f}{\GEV{12}}{-1},
\eeq
where $m_{3/2}$ denotes the gravitino mass, and $M_P \simeq 2.4 \times 10^{18}$\,GeV is
the reduced Planck mass. In the following we assume that $f$ is constant during and after inflation.
We will come back to this issue in Sec.~\ref{sec4}.

Let us assume that the R symmetry is already spontaneously broken during inflation.
To this end we require that inflation scale satisfy the following inequality:
\beq
H_{\rm inf} \;\lesssim\; \GEV{12} \lrf{f}{\GEV{12}}, \label{H<L}
\eeq
where $H_{\rm inf}$ denotes the Hubble scale during inflation. In most part of this letter,
we assume $f \sim \Lambda \sim \sqrt{m_{3/2} M_P}$, although we will show a constraint
in the case of $f \gg \Lambda$~\cite{Izawa:2009mj} later. 
For the R-axion to develop quantum fluctuation that extends beyond the
horizon, the R-axion mass must be lighter than the Hubble parameter during inflation:
\beq
m_a \;\lesssim \; H_{\rm inf}.  \label{ma<H}
\eeq
We focus on the parameter space where the above inequalities are satisfied. 
Indeed, as long as the gravity or anomaly mediation~\cite{Randall:1998uk} is considered, the above
conditions  (\ref{H<L}) and (\ref{ma<H}) are met for plentiful inflation models.

Let us now estimate the R-axion abundance. The R-axion starts to oscillate about
the potential minimum when the Hubble parameter becomes comparable
to its mass, $H \simeq m_a$.
The number density of R-axion $n_a$ devided by the entropy density $s$ is given by
\beq
\frac{n_a}{s} \;\simeq\; \frac{1}{8} \lrfp{a_*}{M_P}{2} \lrf{T_{\rm R}}{m_a},
\eeq
where $T_{\rm R}$ is the reheating temperature after inflation. We have assumed that
the R-axion starts oscillate before the reheating is completed, since otherwise too many
gravitinos would be generated by thermal scatterings (see \EQ{TP}).
Here we have defined an effective initial displacement,
\beq
a_*\equiv{\rm max}~\left\{ f\theta_i, \frac{H_{\rm inf}}{2\pi} \right\},
\eeq
where $\theta_i (= 0 \sim \pi/2)$ denotes the (dimensionless) initial misalignment of the R-axion from its potential minimum.

The decay of the R-axion depends on the SUSY breaking mediation mechanism.
In the gauge mediation~\cite{Giudice:1998bp}, 
since the SUSY breaking field is more strongly coupled to the
visible sector,  the R-axion tends to mainly decay into the supersymmetric standard-model (SSM) 
particles~\cite{Ibe:2006rc,Goh:2008xz}.
On the other hand, in the gravity and anomaly mediation~\cite{Randall:1998uk}, 
we expect that the R-axion  decays into a 
pair of the gravitinos with a non-negligible branching ratio. 
We consider the anomaly mediation in this letter, 
since the two conditions (\ref{H<L}) and (\ref{ma<H}) are satisfied for most of the known inflation models
and  the R-axion will mainly decay into the gravitinos which makes our analysis simple and robust.

The abundance of the gravitino produced from the R-axion decay is
\beq
Y_{3/2} \;\equiv\;\frac{n_{3/2}}{s} = 2B_{3/2} \left(\frac{n_a}{s}\right),
\eeq
where $B_{3/2}$ is the branching fraction of the R-axion decay into gravitinos.
We have
\bea
Y_{3/2}^{(\rm a)} &\simeq&
  3\times 10^{-13} B_{3/2} \theta_i^2 \,
  \lrfp{m_{3/2}}{100{\rm\,TeV}}{-1}
  \lrf{T_{\rm R}}{10^9{\rm\,GeV}}
  \lrfp{f}{10^{12}{\rm\,GeV}}{\frac{5}{2}},
\eea
for $f\theta_i > H_{\rm inf}/(2\pi)$, and
\bea
Y_{3/2}^{(\rm a)} &\simeq&
  7 \times 10^{-19} B_{3/2} \,
  \lrfp{m_{3/2}}{100{\rm\,TeV}}{-1}
  \lrf{T_{\rm R}}{10^9{\rm\,GeV}}
  \lrfp{f}{10^{12}{\rm\,GeV}}{\frac{1}{2}}
  \lrfp{H_{\rm inf}}{\GEV{10}}{2},
\eea
for $f\theta_i < H_{\rm inf}/(2\pi)$, respectively.
On the other hand, the gravitino is also produced by scattering of particles in thermal plasma
and the abundance is given by~\cite{Bolz:2000fu}
\bea
	Y_{3/2}^{(\rm TP)} &\simeq& 2\times 10^{-13}
	\left (1+ \frac{m_{\tilde g}^2}{3m_{3/2}^2} \right )
	\left ( \frac{T_{\rm R}}{10^9~{\rm GeV}} \right ),
	\label{TP}
\eea
where $m_{\tilde g}$ denotes the gluino mass.
Thus the gravtino abundance produced by the R-axion decay can be comparable with 
that from thermal production for $B_{3/2} \sim \theta_i\sim 1$.

In the anomaly-mediation,
the gravitino is so heavy that it decays into the SSM particles
well before BBN. Therefore there is no BBN constraint on the gravitino abundance. 
However, the lightest supersymmetric particle (LSP), which we 
assume to be the Wino-like neutralino, is produced by the gravitino decay. We
obtain the upper bound on the reheating temperature, by requiring that
the neutralino abundance should not exceed the observed cold dark matter (CDM) abundance:
\beq
T_{\rm R} \;\lesssim\; 8 \times 10^9{\rm\,GeV} \lrfp{m_{\tilde W}}{300{\rm\,GeV}}{-1},
\eeq
where $m_{\tilde W}$ is the Wino mass and the gravitino abundance (\ref{TP}) is assumed.\footnote{
	The thermal relic abundance of the Wino-like LSP is smaller than
	the CDM abundance for $m_{\tilde W}\lesssim 3$~TeV~\cite{Hisano:2006nn}.
}
As we will see below, an additional constraint on the R-axion cosmology is obtained by taking account of 
the quantum fluctuations of the R-axion.
In particular, the Hubble scale during inflation is constrained from above by
the WMAP data, even if the abundance of the neutralino originated from
the R-axion is a subdominant component of the dark matter.


\section{Isocurvature Perturbation and Non-Gaussianity\\ from R-axion} \label{sec3}
If the R symmetry is spontaneously broken during inflation with the condition (\ref{ma<H})  satisfied,
the R-axion obtains quantum fluctuations.
Since the energy density of the R-axion is negligibly small compared with that of the inflaton,
its fluctuation can be regarded as an isocurvature perturbation.
This becomes fluctuation in the R-axion abundance after it begins to oscillate.
The R-axion decays into gravitinos, which eventually decay into the SSM particles
producing the neutralino LSP.
Since some fraction of the relic dark matter comes from the R-axion,
this gives rise to the CDM isocurvature perturbation.

The CDM isocurvature perturbation in this model is given by
\beq
	S_{\rm m} = r\left[ 
		\frac{2a_i \delta a}{a_*^2} + \left ( \frac{\delta a}{a_*} \right )^2
	\right],  \label{Sm}
\eeq
where $a_i = f\theta_i$ is the classical deviation from the potential minimum,
$\delta a$ is the quantum fluctuation of the R-axion,
whose Fourier modes satisfy the following condition,
\beq
	\langle \delta a_{\vec k_1}\delta a_{\vec k_2}  \rangle = (2\pi)^3\delta(\vec{k_1}+\vec{k_2})
	\frac{H_{\rm inf}^2}{2k_1^3},
\eeq
 and $r$ denotes the fraction of the DM produced by the
R-axion decay  to the total DM abundance.
The second term in Eq.~(\ref{Sm}) represents the non-Gaussian fluctuation,
and hence, non-Gaussianity resides in the CDM isocurvature perturbation in this model.
The effects of such an isocurvature-type non-Gaussianity was recently studied in detail by
Refs.~\cite{Kawasaki:2008sn,Kawasaki:2008jy,Langlois:2008vk,Moroi:2008nn}
(see also Refs.~\cite{Linde:1996gt,Bartolo:2001cw,Boubekeur:2005fj} for early attempts).
The WMAP5 constraint on the (uncorrelated) isocurvature perturbation reads 
$P_S/P_\zeta < 0.19$ at 95\% C.L. where $P_S$ and $P_\zeta$ are power spectra of
isocurvature and curvature perturbations at a pivot scale $k_0=0.002~{\rm Mpc}^{-1}$
\cite{Komatsu:2008hk}.

Now let us estimate the non-Gaussiniaty.
First, the curvature perturbation $\zeta$ can be expanded as follows according to the
$\delta N$ formalism~\cite{Starobinsky:1986fxa,Sasaki:1995aw,Lyth:2004gb},
\beq
	\zeta = N_a \delta \phi^a + \frac{1}{2}N_{ab}\delta \phi^a \delta \phi^b + \cdots,
\eeq
where $\phi^a$ represent scalar fields which is light during inflation
including the inflaton itself, and $N_a=\partial N/\partial \phi^a$ and so on, with
$N$ representing the local $e$-folding number measured along the world line $\vec x =$const.
In a similar way, the CDM isocurvature perturbation can be expanded as
\beq
	S_{\rm m} = S_a \delta \phi^a + \frac{1}{2}S_{ab}\delta \phi^a \delta \phi^b + \cdots,
\eeq
where $S_a=\partial S_{\rm m}/\partial \phi^a$ and so on.
Then the non-linearity parameter of the isocurvature type, $f_{\rm NL}^{(\rm iso)}$,
is given by~\cite{Kawasaki:2008sn}
\beq
	\frac{6}{5}f_{\rm NL}^{(\rm iso)} = \frac{1}{27}
	\frac{S_aS_bS_{ab}+S_{ab}S_{bc}S_{ca}\Delta_{\delta \phi}^2\ln(k_bL)}{(N_a N_a)^2},
\eeq
where we have introduced an infrared cutoff $L$ taken to be the 
present Hubble scale~\cite{Lyth:1991ub},
and $\Delta_{\delta \phi}^2 = (H_{\rm inf}/2\pi)^2$.
In the present model, there exist two light scalar fields : the inflaton and R-axion.
Since the former is responsible for the curvature perturbation 
and the latter for the isocurvature perturbation,
the non-linearity parameter $f_{\rm NL}^{\rm (iso)}$ is calculated using Eq.~(\ref{Sm}) and
the WMAP normalization condition $|\zeta| \simeq 5\times 10^{-5}$~\cite{Komatsu:2008hk}.
Numerically, it is evaluated as
\beq
	f_{\rm NL}^{(\rm iso)} \sim 2\times 10^{3}~B_{3/2}^3\theta_i^2
	\left ( \frac{T_{\rm R}}{10^9~{\rm GeV}} \right )^3
	\left ( \frac{f}{10^{12}~{\rm GeV}} \right )^{7/2}
	\left ( \frac{H_{\rm inf}}{10^{10}~{\rm GeV}} \right )^4,
\eeq
for $f\theta_i \gtrsim H_{\rm inf}/(2\pi)$, and
\beq
	f_{\rm NL}^{(\rm iso)} \sim 3\times 10^{-3} ~B_{3/2}^3
	\left ( \frac{T_{\rm R}}{10^9~{\rm GeV}} \right )^3
	\left ( \frac{f}{10^{12}~{\rm GeV}} \right )^{3/2}
	\left ( \frac{H_{\rm inf}}{10^{10}~{\rm GeV}} \right )^6.
\eeq
for $f\theta_i \lesssim H_{\rm inf}/(2\pi)$,
where we have used the relation $m_{\tilde W} = (g^2/16\pi^2)m_{3/2}$ 
in the anomaly-mediation with $g$ being the SU(2) gauge coupling constant.

\begin{figure}[ht]
 \begin{center}
   \includegraphics[width=0.5\linewidth]{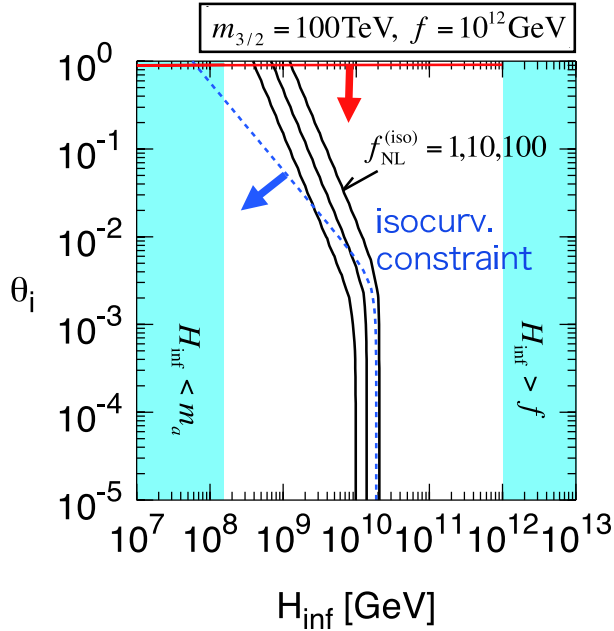} 
   \caption{
  	Contours of $f_{\rm NL}^{(\rm iso)}=1,10$ and $100$  are shown by black solid lines
	from left to right.
	Below the blue dashed line, the isocurvature constraint is satisfied.
	In the shaded regions neither isocurvature perturbation nor non-Gaussianity will arise
	since R symmetry may be restored during inflation 
	($H_{\rm inf}>f$) or R axion is heavier than the Hubble scale during inflation
	($H_{\rm inf}<m_a$).
	Above the red solid line, the neutralino LSP is overproduced.
	In this figure we have taken $f=10^{12}$~GeV.
   }
   \label{fig:f1e12}
 \end{center}
\end{figure}
\begin{figure}[ht]
 \begin{center}
   \includegraphics[width=0.5\linewidth]{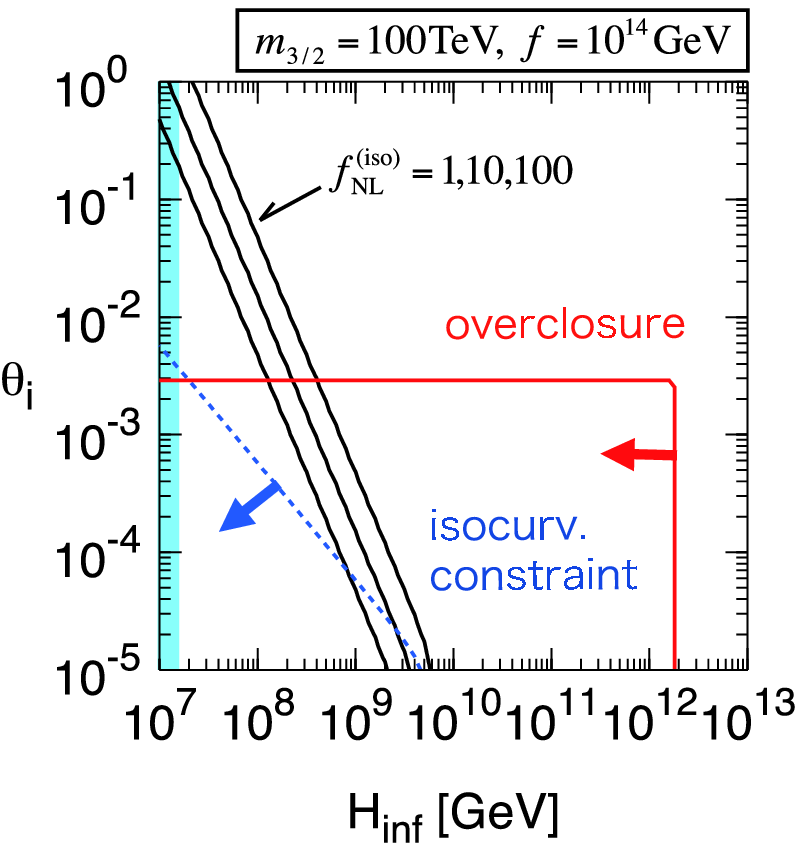} 
   \caption{
  	Same as Fig.~\ref{fig:f1e12}, but for $f=10^{14}$~GeV.
   }
   \label{fig:f1e14}
 \end{center}
\end{figure}

We have shown the constraints on the inflation scale and the initial misalignment 
in Fig.~\ref{fig:f1e12} for $f=10^{12}$~GeV and in Fig.~\ref{fig:f1e14} for $f=10^{14}$~GeV. 
In both figures we have set $m_{3/2}=100$~TeV and 
the reheating temperature $T_{\rm R}$ is taken to be $\sim 8\times 10^9$~GeV
so that the Wino LSP from the gravitinos produced by thermal scatterings 
account for  the observed CDM abundance. 
Below the blue dashed line, the isocurvature constraint is satisfied.
In the shaded regions neither isocurvature perturbation nor non-Gaussianity will arise
since R symmetry may be restored during inflation 
($H_{\rm inf}>f$) or R axion is heavier than the Hubble scale during inflation
($H_{\rm inf}<m_a$).
In these region, our arguments are not applied.
Above the red solid line, the neutralino LSP is overproduced by the decay of 
R-axion-induced gravitino.
In these figures contours of $f_{\rm NL}^{\rm (iso)}=1,10$ and 100 (from left to right) are plotted.
Ref.~\cite{Hikage:2008sk} put a limit on $f_{\rm NL}^{\rm (iso)}$ as
$f_{\rm NL}^{\rm (iso)} < 15$ using Minkowski functional method, and hence
the region with $f_{\rm NL}^{\rm (iso)} \gg 15$ is disfavored.

It is seen that even if the initial misalignment is chosen so as to suppress the R-axion abundance,
the constraints from isocurvature and/or non-Gaussianity exclude large parameter region
once the quantum fluctuation of the R-axion is taken into account.
Remarkably, high scale inflation models with $H_{\rm inf} = 10^{10} \sim 10^{12}$\,GeV are disfavored.


\section{Discussion and Conclusions} \label{sec4}
We have so far assumed that the $U(1)_R$ breaking scale $f$  is constant
during and after inflation. This is the case if the bosonic partner of the R-axion, namely
R-saxion $s$, is stabilized at $\sim f$ during inflation. We are concerned with a class of
SUSY breaking models in which the SUSY breaking field is stabilized 
without taking account of supergravity effects. Therefore, the R-saxion is
heavier than the R-axion, and its mass is expected to be of $\mathcal{O}(m_{3/2} M_P/f)$.
Therefore, for an inflation models with $H_{\rm inf}$ smaller than $\mathcal O(m_{3/2} M_P/f)$,
this assumption is justified. Otherwise, the R-saxion potential generically receives sizable corrections during inflation, 
and the R-saxion may be deviated from the true minimum.
This can affect our analysis in two ways.
First, the isocurvature perturbation is modified. For a larger value of $\la s \ra$,
the isocurvature perturbation becomes smaller and the constraint gets relaxed correspondingly,
where $\la s \ra$ is the expectation value of $|s|$ during inflation.
In particular, the Gaussian part of the isocurvature perturbations are suppressed  by $f/\la s \ra$.   Second, 
the R-saxion will start coherent oscillations, and mainly decays into two R-axions. As long as
the oscillation amplitude is of $\mathcal O(f)$, the produced R-axions do not affect our analysis on the isocurvature perturbation. 
This is not only because the R-saxion is heavier and the number density is suppressed,
but also because, in contrast to the R-axion,  the quantum fluctuations  of the R-saxion can be suppressed due to 
the Hubble-induced mass term.

 We have studied the cosmological impact of the R-axion which arises
 as a result of the spontaneous breaking of the R symmetry in a DSB model.
 The initial misalignment during inflation gives rise to a coherent background
R-axion that decays into gravitinos. In the anomaly mediation,  the gravitino of mass 
heavier than ${\cal O}(10)$\,TeV decays before BBN, and so, 
the R-axion seems cosmologically harmless in this framework. However,
the R-axion acquires quantum fluctuations during inflation if the R symmetry
is spontaneously broken during inflation. Consequently, the Wino LSP DM
has an isocurvature fluctuation with potentially large non-Gaussianity. We have
derived a constraint on the inflation scale and the initial misalignment 
to satisfy the current observations. Our main conclusion is that the R-axion
becomes cosmologically harmless for a sufficiently low inflation scale.

\section*{Acknowledgment}

K.N. would like to thank the Japan Society for the Promotion of
Science for financial support.  The work of F.T. was supported by
JSPS (21740160). This work was supported by 
World Premier International Center Initiative (WPI Program), MEXT,
Japan.

{}

\end{document}